\documentclass[aps,prl,longbibliography,twocolumn,superscriptaddress]{revtex4-2}
\usepackage{amsmath,amssymb,bm,graphicx,color,bbold,hyperref,keyval,url,latexsym,dsfont}
\usepackage{xcolor}
\usepackage[normalem]{ulem} 
\usepackage{CJK}
\usepackage{float}
\usepackage{comment}
\newcommand{\blue}[1]{{\color{black} #1}}
\newcommand{\red}[1]{{\color{black} #1}}

\begin{document}
\title{Density-matrix-renormalization-group-based downfolding of the three-band Hubbard model: the importance of density-assisted hopping}

\begin{CJK*}{UTF8}{}
\author{Shengtao Jiang (\CJKfamily{gbsn}蒋晟韬)}
\altaffiliation{shengtaj@uci.edu}
\affiliation{Department of Physics and Astronomy, University of California, Irvine, California 92697, USA}

\author{Douglas J. Scalapino}
\affiliation{Department of Physics, University of California, Santa Barbara, California 93106, USA}

\author{Steven R. White}
\affiliation{Department of Physics and Astronomy, University of California, Irvine, California 92697, USA}
\date{\today}

\begin{abstract}
Typical Wannier-function downfolding starts with a mean-field or density functional set of bands to construct the Wannier functions. Here we carry out a controlled approach, using DMRG-computed natural orbital bands, to downfold the three-band Hubbard model to an effective single band model. A sharp drop-off in the natural orbital occupancy at the edge of the first band provides a clear justification for a single-band model. Constructing Wannier functions from the first band, we compute all possible two-particle terms and retain those with significant magnitude. The resulting single-band model includes two-site density-assisted hopping terms with $t_n \sim 0.6 t$. These terms lead to a reduction of the ratio $U/t_{\rm eff}$, and are important in capturing the doping-asymmetric carrier mobility, as well as in enhancing the pairing in a single-band model for the hole-doped cuprates. 

\end{abstract}
\maketitle
\end{CJK*}

{\it Introduction.}---
What is the minimal model that captures the important physics of the high-temperature cuprate superconductors? This has been a central question ever since the discovery of the cuprates.
It has been argued that the single-band Hubbard and $t$-$J$ models, in their simplest forms, are sufficient to describe the physics of high $T_{c}$ superconductivity.  
Unexpectedly, recent numerical simulations find that superconductivity \blue{in the ground state} of these single-band models appears to be quite delicate. 
\blue{For example, in the pure Hubbard and $t$-$J$ models ($t',t''=0$),   superconductivity is found to be absent~\cite{absence-qin,tt'j}. 
While the presence of a $t'>0$ can induce superconductivity~\cite{tt'j,sheng2021,jiang2021high,8leg-tj-sl,lu2023,jiang2023hub},  this corresponds to electron doping and the question regarding the presence of hole-doped superconductivity ($t'<0$) is not completely resolved \cite{tt'j,jiang2021high,lu2023,jiang2023hub,dmrg-afqmc-tt'U}.}
The greatest delicacy appears to be associated with the superconductivity;  
other aspects of the models, including antiferromagnetism(AFM) as well as intertwined spin and charge order, appear to be in qualitative agreement with the cuprates~\cite{huang2022intertwined,mai2022intertwined,tt'j,t'',hubreview-qin,tiltstripe,4leghub-huang,karakuzu2022stripe}.

This subtleness of pairing in the single-band models calls for a re-examination of the downfolding process used to derive them, since modest errors could have significant effects. This downfolding is a two-step process, where first one constructs from density functional methods the intermediate-level three-band Hubbard (or Emery) model~\cite{Emery1987}, which includes Cu $d_{x^2-y^2}$, O $p_x$ and O $p_y$ orbitals.  
Since the three-band model is closer to an all-electron Hamiltonian of the cuprates, one expects it to be more reliable than a one-band model---but also more difficult to simulate. There is evidence that the three-band model captures various aspects of the cuprates, particularly magnetic and charge density wave properties ~\cite{huang2017science,mai2021,cui2020,shiwei20,ws2015,dmft3band,threeband-Spalek,lishaozhi}, with greater uncertainty about the pairing properties.
To downfold to a single band model, Zhang and Rice argued that holes on oxygen sites bind to holes on copper sites to form local singlets~\cite{zhangrice}. The Zhang-Rice singlet picture has gained support from experiments~\cite{brookes2001detection,tjeng1997spin,harada2002unique} as well as calculations~\cite{cui2020,mai2021,arrigoni2009phase,Aligia1}, and motivated studies of various single-band Hubbard~\cite{vmc2021,4leghub-yfjiang,stripe-vmc1,stripe-vmc2,stripe-afqmc,hubstp-Ido,hubstp-tocchio,senechal2005,machida1,xu2020competing,himeda2002stripe,jiang2022stripe,xu2022stripes,hub-bench2015,stripehubbard,Peng-attraction,Wietek-hubbard,weak-coupling-sc,boxiao-hubbard,fourbyfour_ed,jiang2023hub} and $t$-$J$ models~\cite{Corboztj2014,4legtj-hcjiang,4legtj-dodaro,corboztj2011,sheng2021,vmctjstripe,yaowang2022,finiteT-tt'j,t'',8leg-tj-sl,twolegtj}. 

Here we demonstrate an alternative way to downfold the three-band Hubbard model based on a density-matrix renormalization group (DMRG)~\cite{dmrg} construction of Cu-centered Wannier functions.
The general idea of constructing effective models using {\it ab initio} calculations has been explored in various contexts~\cite{chang-wagner,changlani-honeycomb,cui-science,changlani,nickelates-downfold}.
Our approach uses DMRG to compute the natural orbitals of the three-band model, and from those construct Wannier functions, similar to a recent work that downfolds hydrogen chains into Hubbard-like models~\cite{randy}.
The resulting single-band model includes additional two-site density-assisted hopping terms $t_n$ whose magnitude is comparable to $t$.  On a mean-field level, these new terms simply reduce the ratio $U/t_{\rm eff}$, \blue{with $t_{\rm eff}=t+t_n\langle n \rangle$, where $\langle n \rangle$ is the average number of holes per CuO$_2$ unit cell}.  However, beyond mean-field, the $t_n$ terms capture the doping-asymmetric carrier mobility, and, as revealed by a measurement of the superconducting phase stiffness, further enhance the pairing in the hole-doped single-band model.

\begin{figure}[t]
\centering    
\includegraphics[width=1.0\columnwidth]{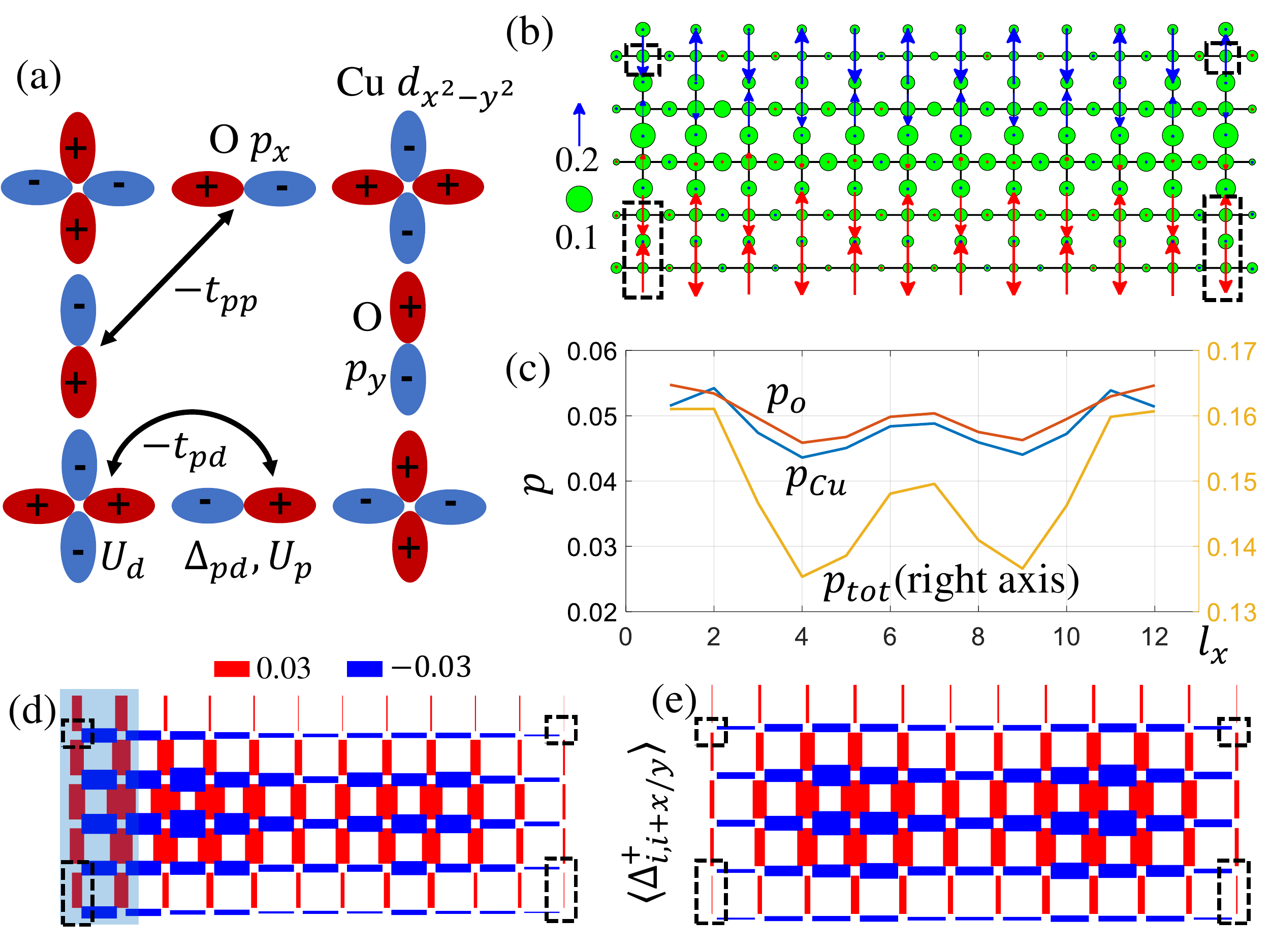}
\vskip -0.25cm
\caption{(a): The three-band Hubbard model and our phase convention for the orbital basis. (b): Charge and spin structure on a $12 \times 5$ cylinder at a hole doping $\sim 0.15$. The length of the arrows and the diameter of the circles represent $\langle S^z \rangle$ and local doping, respectively. The spins are colored to indicate different AFM domains. There are weak magnetic pinning fields applied on the boundary sites in the dotted boxes. (c): Average orbital-resolved local doping $p_{\rm Cu/O}$ along the length of the cylinder \blue{with $p_{\rm tot}$=$p_{\rm Cu}+2p_{\rm O}$} (d) and (e): Pairing order $\langle \Delta^\dagger_{ij} + \Delta_{ij} \rangle$ between neighboring Cu sites $i$ and $j$. The thickness/color of the bond indicates the magnitude/sign of the pairing. The pairing orders away from the edges are similar for (d) which has pairfields applied on the shaded left edge and (e) which spontaneously breaks symmetry.
}
\label{fig:threeband}
\vspace{-0.5cm}
\end{figure}

{\it The three-band model.}---We present the lattice structure and the terms in the three-band Hubbard model in Fig.~\ref{fig:threeband}(a). Each CuO$_2$ unit cell consists of three orbitals: Cu $d_{x^2-y^2}$, O $p_x$ and O $p_y$. \blue{We study clusters with cylindrical boundary conditions. For an $L_x$ by $L_y$ cylinder, there are $N_{\rm Cu}$=$L_xL_y$ Cu sites and $N_{\rm O}$=$(2L_x+1)L_y$ O sites.} In the undoped insulator at half-filling, there is one hole per unit cell, and the model is written in the hole picture with $d^\dagger_{i\sigma}$ or $p^\dagger_{j\sigma}$ creating a hole with spin $\sigma$ on a Cu site $i$ or O site $j$. 
Hole doping corresponds to $\langle n \rangle>1$ while electron doping corresponds to $\langle n \rangle<1$.
The three-band Hamiltonian is:
\begin{equation}
\begin{aligned}
    \label{eq:threeband}
    H^{TB}&=-t_{pd}\sum_{\langle ij \rangle \sigma}(d^\dag_{i\sigma}p_{j\sigma}+h.c.)
    -t_{pp}\sum_{\langle \langle ij \rangle \rangle \sigma}(p^\dag_{i\sigma}p_{j\sigma}+h.c.)\\
    &+U_{d}\sum_{i}n^d_{i\uparrow}n^d_{i\downarrow}
    +U_{p}\sum_{i}n^p_{i\uparrow}n^p_{i\downarrow}
    +\Delta_{pd}\sum_{i \sigma}p^\dag_{i\sigma}p_{i\sigma}
\end{aligned}
\end{equation}
where $t_{pd}$/$t_{pp}$ hops a hole between nearest-neighbor Cu-O/O-O sites, and \blue{the summation $\langle ij \rangle$/$\langle \langle ij \rangle \rangle$} runs over all relevant pairs of sites. We have chosen a gauge \blue{for the orbitals as shown in Fig.~\ref{fig:threeband}(a)} so that all hoppings are negative; $U_d$ and $U_p$ are the on-site repulsion term on the Cu and O sites; $\Delta_{pd}$=$\epsilon_p-\epsilon_d$ is the energy difference for occupying an O site compared to occupying a Cu site. We set the energy scale with $t_{pd}=1.0$, and take  $t_{pp}=0.5$, $U_d=6.0$, $U_p=3.0$, and $\Delta_{pd}=3.5$, unless otherwise noted, which appropriately describes a charge-transfer system where $U_d>\Delta_{pd}$ and $\Delta_{pd}>2t_{pd}$.
\blue{Estimates for $t_{pd}$ range from 1.1eV~\cite{1.1ev} to 1.5eV~\cite{hanke20103}.}
Comparing with previously used parameters~\cite{ws2015,hanke20103}, here we increase $\Delta_{pd}$ to incorporate the effect of $V_{pd}$, and choose a somewhat smaller $U_d$ 
for a stronger pairing response~\cite{sm}. \blue{Systems \textbf{h1} and \textbf{e1} have hole and electron dopings of 0.15.}  Another hole-doped case \textbf{h2} with $U_d$=3.5 and $\Delta_{pd}$=5.0  describes a Mott-Hubbard rather than charge-transfer system~\footnote{See Ref.\cite{ct-mh} for detailed definitions of a charge-transfer insulator and Mott-Hubbard insulator.}.
The calculations are carried out using the $\mathtt{ITensor}$ library~\cite{itensor}. We typically perform around 20 sweeps and keep a maximum bond dimension of ~7000 to ensure convergence with a maximum truncation error of $\mathcal{O}(10^{-5})$.

Previous studies of the three-band model have identified features consistent with the cuprates, including doping asymmetry, formation of stripes on the hole doped side and commensurate AFM on the electron-doped side~\cite{ws2015,shiwei20}. There is evidence for $d$-wave pairing  for both electron and hole doping, with the dominant component between nearest neighbor Cu sites~\cite{cui2020,mai2021,threeband-Spalek,pdw-hcjiang,tblad-song}.

Of particular concern for finite size effects is the quantization of stripe filling around a short cylinder~\cite{ws2015}; here we choose a width-5 cylinder so that one stripe can form lengthwise; see Fig.~\ref{fig:threeband}(b).  The Cu-Cu pairing is shown in Fig.~\ref{fig:threeband}(e). Along the stripe an additional pairing modulation reflects an edge-induced charge density oscillation, as shown in Fig.~\ref{fig:threeband}(c). 
\red{Similar pairing occurs whether it is pinned by edge pair fields [Fig.~\ref{fig:threeband}(d)] or allowed to arise spontaneously as a finite bond dimension broken symmetry~\cite{tt'j} [Fig.~\ref{fig:threeband}(e)]}. \blue{The existance of pairing for a hole-doped three-band model has also been reported in a recent infinite projected entangled-pair states study~\cite{ipeps}.}

\begin{figure}[t]
\centering    
\includegraphics[width=1.0\columnwidth]{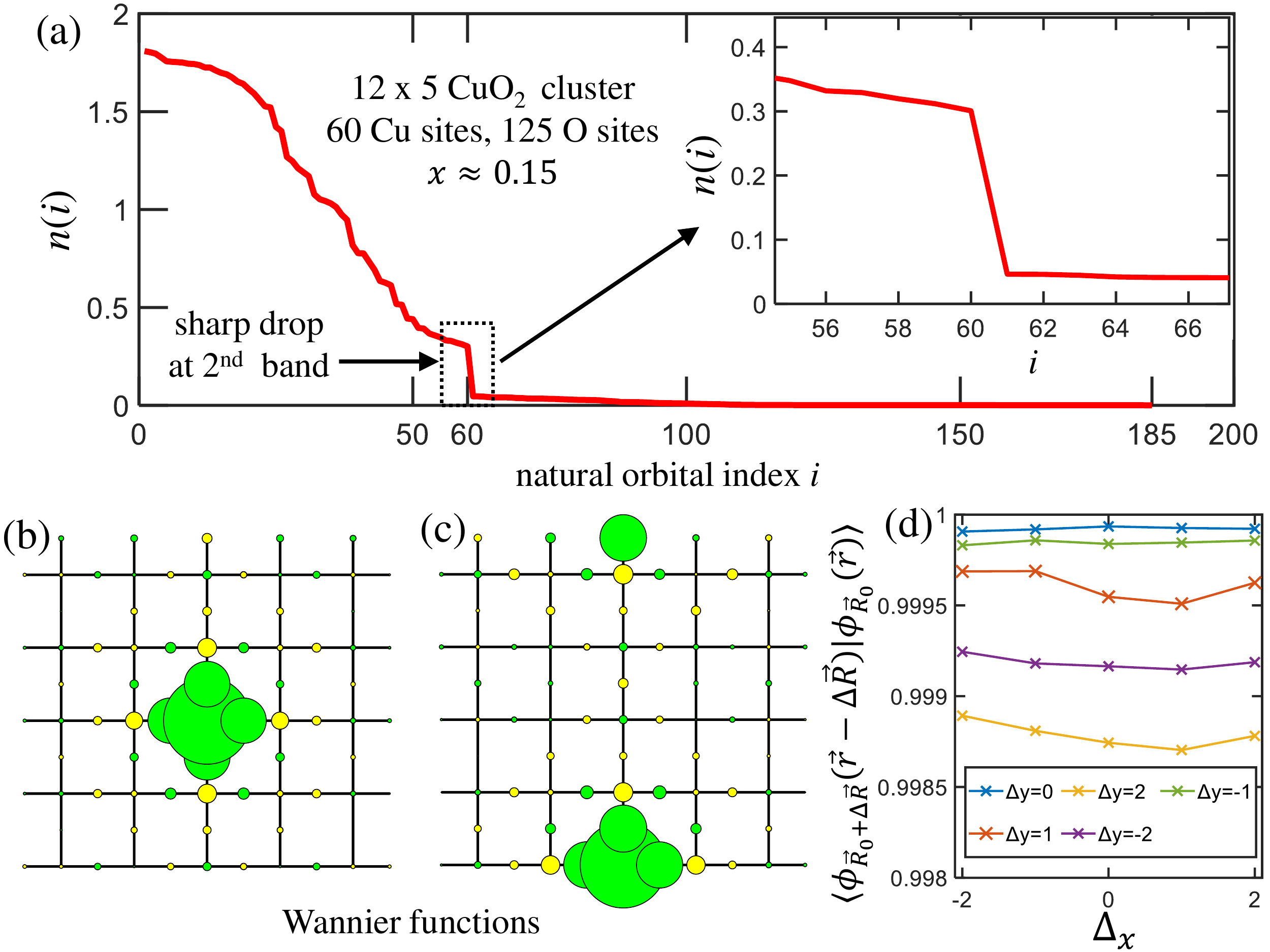}
\vskip -0.25cm
\caption{At a hole doping of 0.15 (a): occupancies of the natural orbitals obtained by diagonalizing the single-particle correlation matrix $M_{\alpha \beta}=\sum_{\sigma}\langle C^\dagger_{\alpha\sigma} C_{\beta\sigma} \rangle$ , with $C^\dagger=\{d^\dagger,p_x^\dagger,p_y^\dagger\}$. The natural orbital states/occupancies correspond to the eigenvectors/eigenvalues of $M_{\alpha \beta}$. The inset is a zoom-in of the region that shows a sharp drop at the second band beyond which occupancies are limited ($<2\%$). (b) and (c): Cu-centered Wannier functions at two different locations constructed from the natural orbitals of the first band. Color/area of the circles indicate the sign/magnitude of the local orbital component. (e): overlap of Wannier functions (truncated to a $5 \times 5$ CuO$_2$ unit cell) with their centers shifted to the same site, showing they are almost translational invariant.}
\label{fig:nat-orb}
\vskip -0.5cm
\end{figure}

\blue{{\it Downfolding into a Wannier single-band model.}---The occupied bands in the DMRG wavefunctions are identified by measuring  the single-particle correlation matrix $M_{\alpha \beta}=\sum_{\sigma}\langle C^\dagger_{\alpha\sigma} C_{\beta\sigma} \rangle$, with $\{C^\dagger\}=\{d^\dagger,p_x^\dagger,p_y^\dagger\}$, whose eigenvectors and eigenvalues define the natural orbitals (NOs) and their occupancies, respectively.
In a non-interacting system, the NO occupancies make a step function at the Fermi level. 
Here, this step near $i \sim 35$ is completely smeared out [Fig.~\ref{fig:nat-orb}(a)], reflecting the strong correlation in the system.
However, there is a sharp drop in occupancies  at $i=60$, the total number of Cu sites, indicating the end of the first band. Beyond the first band, the total occupancy is $<2\%$, and for the electron doped case, $<0.4\%$. This provides a strong justification for downfolding into a single-band, which would be exact if the higher-band occupancies were zero.
We observe similar sharp drop-offs for narrower width 2 and 4 systems.  This indicates that the drop-off is due to short-range physics involving the Cu and surrounding O orbitals, which can be seen clearly on small systems.
We observe a similar sharp drop-off for a range of three-band parameters, including in the Mott-Hubbard regime.  

Given the accuracy of the truncation to a single band, we can derive an effective single-band model through the standard Wannier construction with a simple single-particle transformation. We first localize the functions of this band into Cu-centered Wannier functions (WFs), see Ref.~\cite{sm} for details. We show two representative WFs in Fig.~\ref{fig:nat-orb}(b) and (c), which are evidently highly localized. Functions on different sites are almost identical;  evidence for this translational invariance is shown in Fig.~\ref{fig:nat-orb}(d).

To construct the effective Hamiltonian in the WF space, we first organize the WFs into a $N_{\rm Cu}$-by-($N_{\rm Cu}+N_{\rm O}$) real isometric matrix $\bf{A}$ ($\bf{AA^\dagger}$=$\mathds{1}$) , with entry $A_{ij}$ being the weight of the three-band orbital $j$ in the Wannier function centered at Cu site $i$~\footnote{The matrix elements of $\bf{A}$ are listed in Ref.~\cite{sm} Sec.II}. The matrix $\bf{A}$ defines a single-particle transformation from the three-band basis $\{C^\dagger\}=\{d^\dagger,p_x^\dagger,p_y^\dagger\}$ to the WF basis $\{c^\dagger\}$:
\begin{equation}
\label{eq:trans}
\begin{aligned}
&c^\dagger_i=\sum_j A_{ij}C^\dagger_j \\
\end{aligned}
\end{equation}
We invert this relationship, taking:
\begin{equation}
\begin{aligned}
\label{eq:inv}
&C^\dagger_j=\sum_i A_{ij}c^\dagger_i+\text{higher bands}  
\end{aligned}
\end{equation}
where we omit the higher bands.
The Wannier Hamiltonian is obtained by inserting Eq.~\ref{eq:inv} into the three-band Hamiltonian [Eq.~\ref{eq:threeband}].
The single-particle terms $k_{\alpha \beta}$, which include the $t_{pd}$, $t_{pp}$ and $\Delta_{pd}$ terms, become 
\begin{equation}
k_{\alpha \beta}C^\dagger_{\alpha\sigma}C_{\beta\sigma} \rightarrow k_{\alpha\beta}\sum_{ij}A_{i\alpha}A_{j\beta}~c_{i\sigma}^\dagger c_{j\sigma}.
\end{equation}
The two-particle terms $U_\alpha$, which include the $U_d$ and $U_p$ terms, become
\begin{equation}
U_\alpha n_{\alpha \uparrow}n_{\alpha \downarrow} \rightarrow U_\alpha \sum_{ijkl}A_{i\alpha}A_{j\alpha}A_{k\alpha}A_{l\alpha}~c_{i\uparrow}^\dagger c_{j\uparrow}~c_{k\downarrow}^\dagger c_{l\downarrow}.
\end{equation}

Although the Wannier Hamiltonian has $\mathcal{O}(N^2)$ single particle and $\mathcal{O}(N^4)$ two particle terms, both the single-particle and two-particle terms decay quickly with the distance between sites.
Magnitudes of the  single-particle hoppings beyond third-nearest neighbors are smaller than $0.01t$ and are truncated. 
The largest two-particle term is the onsite repulsion $U$. The second largest is the nearest-neighbor density-assisted hopping $t_nc^\dagger_{j,\sigma} c_{i,\sigma} n_{i\Bar{\sigma}}$.
We also keep the second and third nearest-neighbor density-assisted hoppings ($t_n'$ and $t_n''$). All other two-particle terms are less than 0.05$t$ and are truncated. After these simplifications, we obtain a {\it truncated Wannier model}:
\begin{equation}
\begin{aligned}
    \label{eq:wannier}   
    &H=\sum_{i,\delta,\sigma}
    -t^{\delta} c^\dagger_{i+\delta,\sigma} c_{i,\sigma}+\sum_{i} Un_{i,\uparrow}n_{i,\downarrow}\\ &+\sum_{i,\delta_i,\sigma}-t^{\delta}_n (c^\dagger_{i+\delta,\sigma} c_{i,\sigma}+c^\dagger_{i,\sigma} c_{i+\delta,\sigma})
    n_{i\Bar{\sigma}}. 
\end{aligned}
\end{equation}
Here $i+\delta$ is the first, second, or third nearest neighbor of site $i$, with conventional hopping amplitudes $t$, $t'$, and $t''$, and with density-assisted hopping amplitudes $t_n$, $t'_n$, $t''_n$. The resulting model parameters are summarized in Table.~\ref{table:wan} for downfolding based on three different three-band systems~\footnote{The hopping coefficients are averaged over horizontal and vertical directions, which typically differ by $\sim 10\%$, but somewhat more (-0.07 and -0.11) in the case of $t''_n$ in system \textbf{h1}.}. 

\begin{table}[h]
\centering
\vskip -0.3cm
\caption{Parameters for the Wannier single-band model from downfolding the three-band model.  \textbf{h} and \textbf{e} correspond to hole and electron doping of 0.15. $t_{pd}$ is nominally 1.5eV.}

\setlength{\tabcolsep}{1.3mm}{
\begin{tabular}[b]{cccccccc}
\hline
\hline
\textbf{case ($U_d,\Delta_{pd}$)} &$t/t_{pd}$ &$t_n/t$ &$t'/t$ &$t'_n/t$ &$t''/t$ &$t''_n/t$ &$U/t$\\
\hline
\textbf{h1} (6.0,~3.5) &0.27 &0.60 &0.07 &0.05 &-0.04 &-0.09 &12.6\\
\textbf{e1} (6.0,~3.5) &0.28 &0.52 &0.08 &0.08 &-0.05 &-0.04 &13.7\\
\textbf{h2} (3.5,~5.0) &0.21 &0.33 &0.08 &0.05 &-0.03 &-0.04 &11.8\\
\hline
\end{tabular}}
\label{table:wan}
\vspace{-0.2cm}
\end{table}

Note that the nearest-neighbor $t_n$ coefficients are almost twice the size of an effective exchange coupling $J\sim 4 t^2/U\sim 0.32$. Given their substantial magnitude, it is surprising how rarely these terms have been considered~\cite{tn_georges,tn_hirsch,tn_Laughlin,tn_sondhi}. The existance of the $t_n$ term is guaranteed by a finite component of the nearest-neighbor Cu orbitals in the Wannier function, which is robust since regular Wannier functions must have those components to satisfy orthogonality. Its magnitude is substantial mainly because of the large value of $U_d$.
The $t_n$ term is much larger for the cuprate-relevant charge-transfer case \textbf{h1}, compared with the Mott-Hubbard case \textbf{h2} that has a similar $U/t$ ratio. This is directly tied to the higher O-occupancy in the charge-transfer case, which makes the WFs more extended.

\begin{figure}[t]
\centering    
\includegraphics[width=1.0\columnwidth]{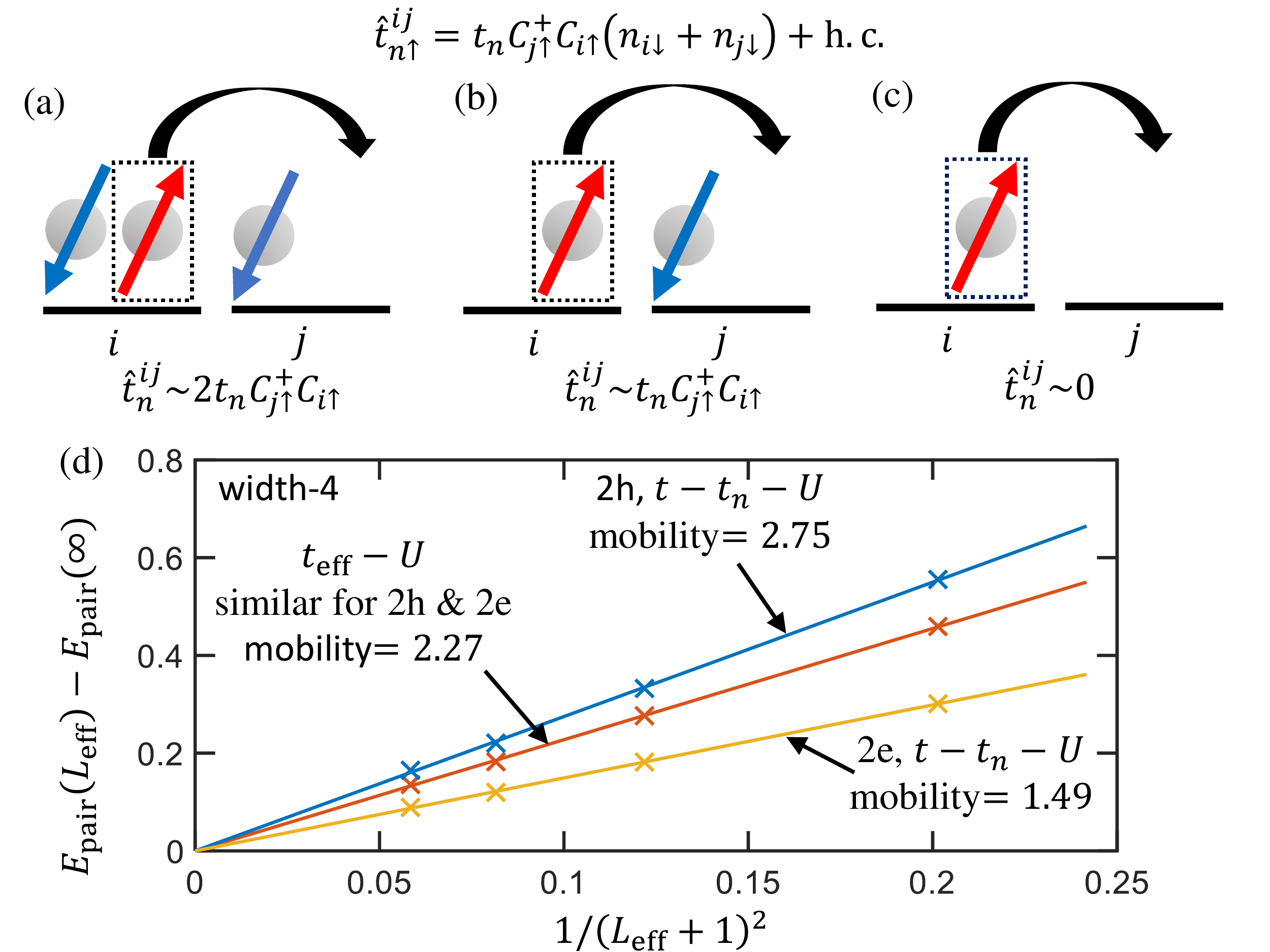}
\vskip -0.2cm
\caption{(a),(b) and (c): action of the $t_n$ term, and resulting hopping strengths, depending on the occupations of the sites involved. (d) On a width-4 cylinder, mobility of a pair of holes/electrons (in unit of $t$) measured by the slope of pair energy versus $1/(L_{\rm eff}+1)^2$, for the $t$-$t_n$-$U$ model and the $t_{\rm eff}$-$U$ model.}
\vspace{-0.5cm}
\label{fig:tn}
\end{figure}

We also note that the WFs and thus the model parameters are similar for the hole and electron doped cases, even if their parental three-band states are quite different in spin and charge order, indicating that the downfolding is determined by the local physics.
Just as the sharp drop in occupancy after the first band shows little dependence on system size, we find the Wannier Hamiltonian also exhibits little dependence on cluster size.

Two key questions now arise: (1) does the Wannier model Hamiltonian give the same properties as the three band model? Given the straightforward and robust nature of our downfolding, we expect this to be so, and comparisons detailed in the Supplementary Material~\cite{sm} for moderate system sizes support this. (2) Does a mean-field treatment of the $t_n$ terms, reducing the system to a standard Hubbard model, also match the properties of the three-band model? Although this may be largely true for the spin and charge degrees of freedom, we will argue that the delicate nature of the pairing is not correctly captured by the mean field/standard Hubbard treatment. In any case, the large magnitude of $t_n$ poses a potential difficulty for a mean-field treatment since any deviations could be significant.}


{\it Effects of $t_n$}---We find that the $t_n$ terms have two primary effects: first, they reduce the effective interaction strength $U/t_{\rm eff}$; and second, they enhance hole hopping, reducing the effective mass of pairs on the hole-doped side and promoting phase coherence.
The reduction of $U/t_{\rm eff}$ can be understood from a mean-field treatment of $t_n$ where one replaces $t_n c^\dagger_{j\sigma}c_{i\sigma}(n_{i\Bar{\sigma}}+n_{j\Bar{\sigma}})$ by $t_nc^\dagger_{j\sigma}c_{i\sigma}\langle n\rangle$, with $\langle n\rangle$ being the average density of holes per Cu site, adding to the conventional hopping.  This changes $U/t \sim 13$ to $U/t_{\rm eff}\sim7.5$ (for $t_n$=0.6, $n$=1.15), close to $U/t=8$, which is often used for the cuprates.

Beyond mean-field, we consider specific hopping processes in Fig.~\ref{fig:tn}(a-c), written in the hole-picture.  
For a doped hole (i.e. a doublon) we expect process (a) to be relevant, where the $t_n$ acts with magnitude $2t_n$. For undoped regions with AF particle-hole virtual hoppings, process (b) acts with magnitude $t_n$. On the  
electron doped side, process (c), $t_n$ has no effect. It does not seem possible to capture these various properties correctly with a mean field treatment.

We find that the resulting hole-pair mobility is enhanced with the $t_nc^\dagger_{j,\sigma} c_{i,\sigma} n_{i\Bar{\sigma}}$ term versus its mean-field $t_nc^\dagger_{j,\sigma} c_{i,\sigma} \langle n_{i\Bar{\sigma}}\rangle$, $2.75t$ versus $2.27t$. In contrast, the mobility of a pair of electrons with $t_n$ is much smaller, 1.49$t$, and reduced comparing to its mean-field $2.27t$.  \blue{Thus, the increased mobility of a single pair hints at the possibility of enhanced pairing due to $t_n$ on the hole doped side. }

\begin{figure}[t]
\centering    
\includegraphics[width=1.0\columnwidth]{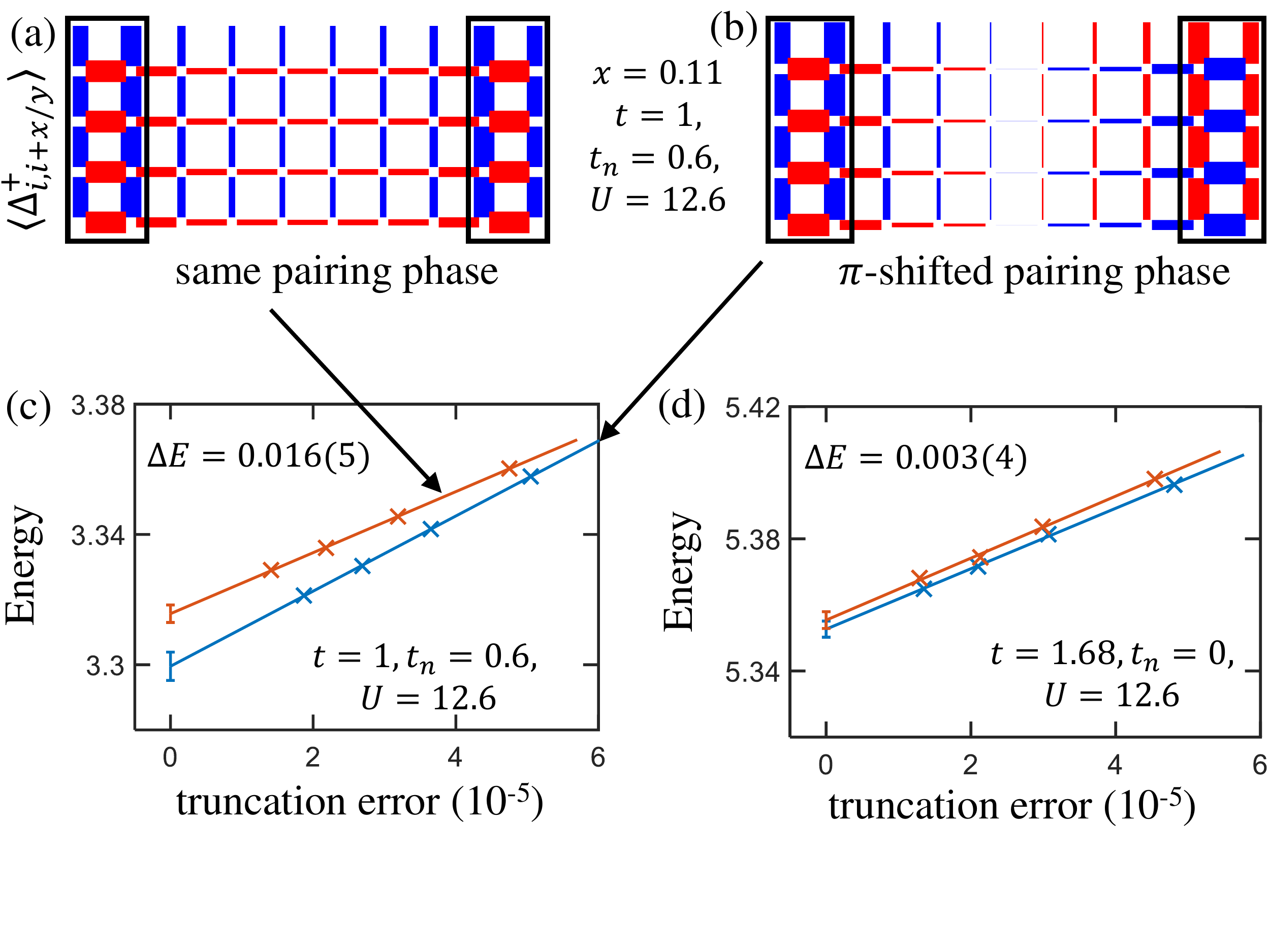}
\vskip -0.95cm
\caption{Pairing response for the $t$-$t_n$-$U$ model in a $10 \times 4$ cylinder at a hole doping $\sim 0.11$($n\approx1.11$). Pair-fields have been applied to regions near both edges, denoted by the black boxes, with the phases on the two ends (a) being the same and (b) having a $\pi$ shift. (c): extrapolation of the energies with the truncation errors for the two different pairfield boundary conditions in (a) and (b). The energy difference is a measurement of the superconducting phase stiffness. (d) Same as (c) for the $t_{\rm eff}$-$U$ model that incorporates the effect of the $t_n$ term only in mean-field.}
\label{fig:phase-stiff}
\vspace{-0.5cm}
\end{figure}

\blue{To probe for superconductivity, we apply edge pairfields to a $10\times4$ cylinder with and without a $\pi$ phase shift between the two edges, to measure the  superconducting phase stiffness $\alpha$.
The results are shown in Fig.~\ref{fig:phase-stiff}.
Note that $\alpha=0$ indicates the absence of superconductivity.  The applied fields make $\alpha$  proportional to an energy difference, $\alpha \propto \frac{L_x}{L_y}\Delta E$, where $\Delta E$ can be extrapolated using DMRG.  At a hole doping of 0.11 ($\langle n\rangle \approx1.11$), the $t$-$t_n$-$U$ model gives a stiffness $\alpha$ that is five times larger than the $t_{\rm eff}$-$U$ model~\footnote{In the supplementary material~\cite{sm}, we vary $t_n$ from 0 to 1.6 and find that the superconducting phase stiffness gets bigger as $t_n$ increases, and the phase stiffness is greater compared to using a mean-field $t_{\rm eff}$.}. 
The pure Hubbard model (without $t'$ terms) is thought to be non-superconducting~\cite{absence-qin}; our results hint that the $t_n$ terms, even without $t'$, might tip the balance towards superconductivity.
In a more realistic model where $t'$ and $t'_n$ from Table.~\ref{table:wan} are included, we also find a larger phase stiffness, $\Delta E= 0.012(4)$ with $t_n$ versus $0.002(4)$ with $t_{\rm eff}$, for a system at a hole doping of 0.11. }

{\it Summary and discussion.}---
We have revisited the Zhang-Rice downfolding of the three-band Hubbard model to a single-band model, basing the downfolding on a DMRG simulation of the three band model. 
Our results give strong support to the applicability of the one band approach, where the small occupancy of higher natural orbital bands shows their irrelevance. However, our Wannier function downfolding also shows that a density-assisted hopping term which is usually neglected has a large coefficient. 
This term renormalizes the hopping in mean field, but mean field treatments are inadequate to capture the effects of this term on pairing.
The density-assisted hopping enhances hole mobility and hole-pair mobility.  This leads to enhanced superconducting pairing on the hole-doped side on width-4 cylinders.


\begin{acknowledgments}
\emph{Acknowledgments.}---
We acknowledge helpful discussion with A. Georges, H.-C. Jiang, S. Sondhi and S. A. Kivelson.
SJ and SRW are supported by the NSF under DMR-2110041.
\end{acknowledgments}

\bibliography{ref}
\end{document}


\title{Density-matrix-renormalization-group-based downfolding of the three-band Hubbard model: the importance of density-assisted hopping: supplemental Materials}

\begin{CJK*}{UTF8}{}
\author{Shengtao Jiang (\CJKfamily{gbsn}蒋晟韬)}
\altaffiliation{shengtaj@uci.edu}
\affiliation{Department of Physics and Astronomy, University of California, Irvine, California 92697, USA}

\author{Douglas J. Scalapino}
\affiliation{Department of Physics, University of California, Santa Barbara, California 93106, USA}

\author{Steven R. White}
\affiliation{Department of Physics and Astronomy, University of California, Irvine, California 92697, USA}
\date{\today}

\maketitle
\end{CJK*}

\section{The parameters choice for the three-band Hubbard model}

\begin{figure}[h]
\centering
\includegraphics[width=0.7\columnwidth]{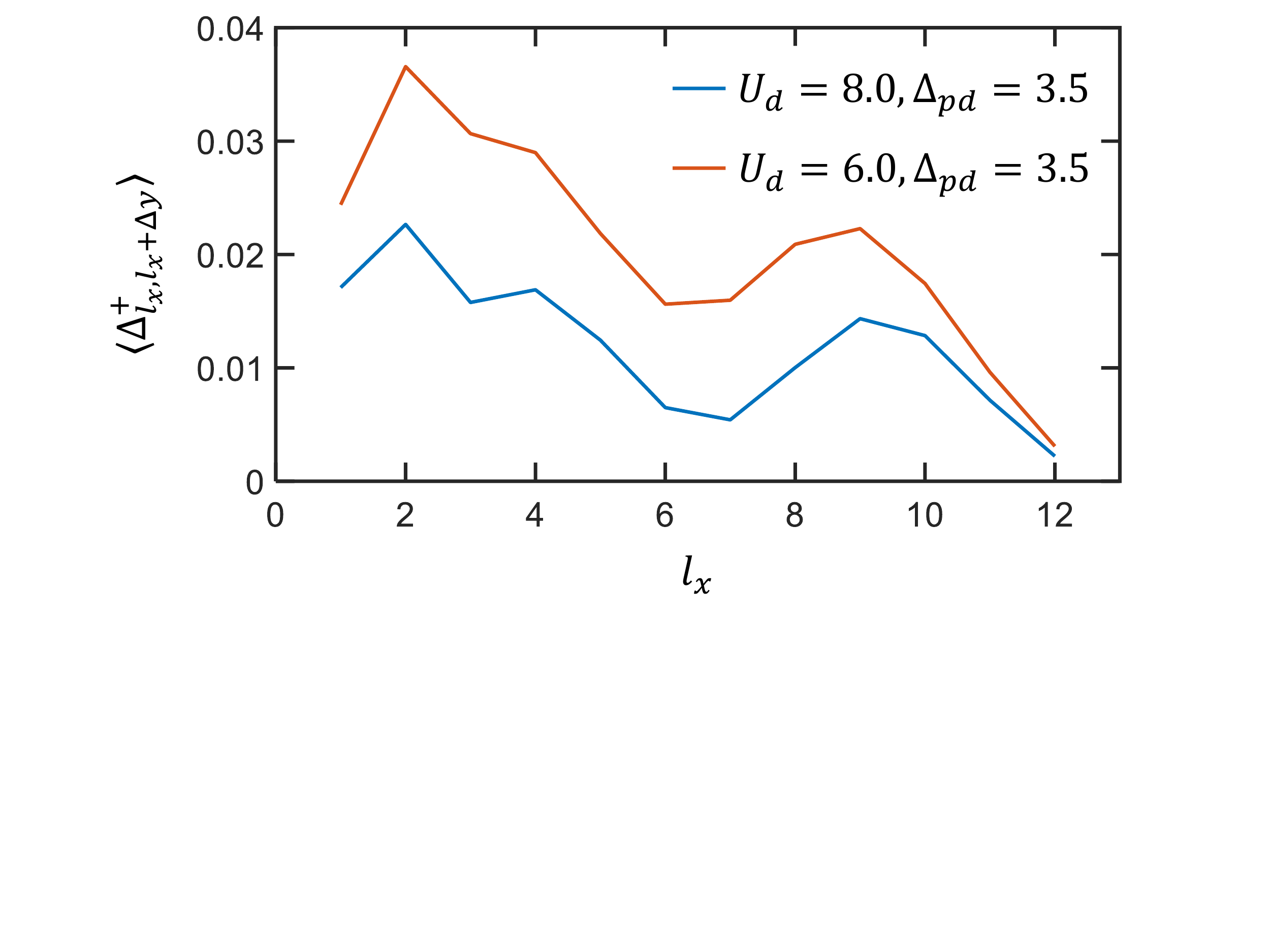}
\vskip -3.7cm
\caption{Pairing response of two different $U_{dd}$ with pairfields applied on two columns ($l_x=1,2$) on the left edge of a 12 $\times$ 5 cylinder, at a hole doping of 0.15 ($n=1.15$). Both the applied pairfields and the pairing responses are on nearest-neighbor vertical Cu-Cu bonds.}
\label{fig:pairing}
\end{figure}

A common choice of the parameters for the three-band Hubbard model is: $t_{pd}=1.0$, $t_{pp}=0.5$, $\Delta_{pd}=3.0$, $U_{d}=8.0$, $U_{p}=3.0$, $V_{pd}=0.5$~\cite{hanke20103}.
In a recent density-matrix embedding theory study~\cite{cui-threeband}, the appropriate range of $U_{d}$ is estimated to be 4.5-9.3 when fixing other parameters as in the common choice. While the common choice $U_{d}=8.0$ falls in the upper half of this range, we also consider another $U_{d}=6.0$ in the lower half of this range.
In addition, we omit $V_{pd}$ but increase $\Delta_{pd}$ to 3.5 to incorporate its effect.
A comparison of the pairing responses of these two choices of $U_{d}$ are shown in Fig.~\ref{fig:pairing}, and $U_{d}=6.0$ exhibits a stronger pairing response, so we choose it for a clearer signal.

\section{Details of constructing Wannier functions}

Our construction of the Wannier functions and subsequent downfolding is patterned after Ref.~\cite{randy}, with the key difference being that in our study we start with a three band model rather than a continuum all electron calculation described by sliced basis functions. 
The goal is to form  Wannier functions centered at Cu site $j$ $\{\phi_j(\Vec{r})\}$ by linearly combining the natural orbitals of the first-band $\{\psi_i(\Vec{r})\}$, with both $i$ and $j$ ranging from 1 to the total number of Cu sites. We would like the Wannier functions to be: (1) orthonormal, (2) translationally invariant, and (3) localized around Cu sites. These properties ensure the downfolded Wannier Hamiltonian has short-ranged interactions with site-independent magnitudes. While the translational invariance cannot be completely achieved because we are dealing with cylindrical systems with open edges, we expect the Wannier functions in the bulk to almost satisfy the above properties.

First we construct functions $\{\delta_j(\Vec{r})\}$ localized around each Cu site $j$ by superposing the natural orbitals $\{\psi_i\}$ with coefficients being their respective weight on Cu site $j$:
\begin{equation}
    \delta_j(\Vec{r})=\sum_i \psi_i(\Vec{r}=\Vec{r}_j) \psi_i(\Vec{r})
\end{equation}
where $\Vec{r}_j$ is the position vector corresponding to Cu site $j$. The functions $\{\delta_j(\Vec{r})\}$ are localized, but not orthonormal. We orthonormalize them while preserving their locality using L{\"o}wdin symmetric orthogonalization~\cite{symortho}, which minimally rotates these functions (note $O$ is a matrix):
\begin{equation}
\begin{split}
\phi_j(\Vec{r})=\sum_j {[O^{-\frac{1}{2}}]}_{jj'}~\delta_{j'}(\Vec{r}), \\
\text{with} \: O_{jj'}=\langle \delta_{j}(\Vec{r})~|~\delta_{j'}(\Vec{r}) \rangle.
\end{split}
\end{equation}

\begin{figure}[h]
\centering    
\includegraphics[width=0.7\columnwidth]{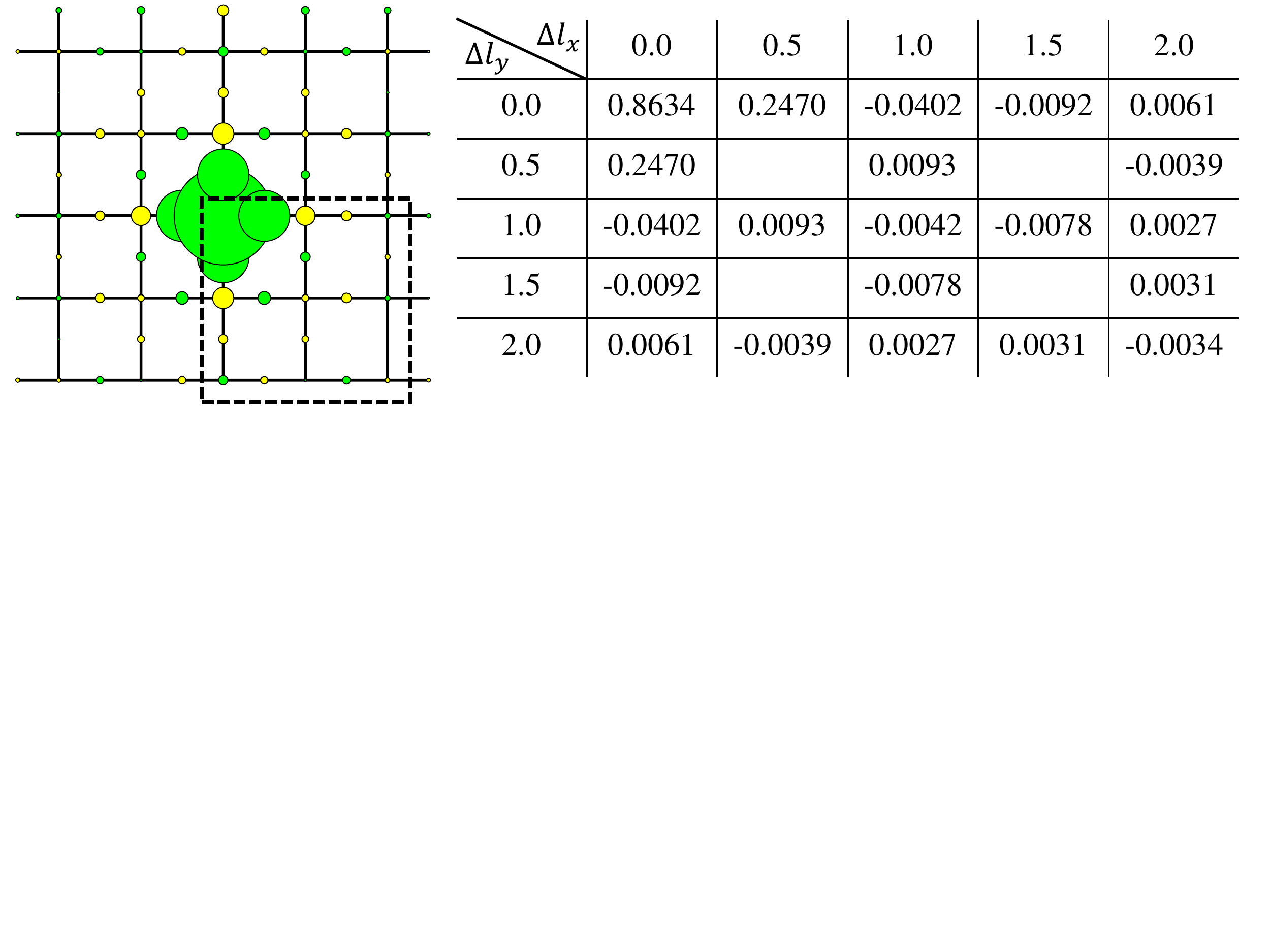}
\vskip -5.5cm
\caption{A Cu-centered Wannier function with the color/area of the circles denoting the sign/magnitude of the local orbital components. The coefficients of the orbitals in the lower-right quadrant are shown in the table, which have been averaged over the vertical and horizontal directions. The Wannier function here is from downfolding the case \textbf{h1} in the main text.}
\label{fig:wan}
\end{figure}

The resulting Wannier functions $\{\phi_j(\Vec{r})\}$ are guaranteed to be orthonormal and are almost translational invariant, as we show in Fig.~2(e) in the main text.  We plot the structure of the resulting Wannier function in Fig.~\ref{fig:wan}. It is localized around one Cu site with its tail decaying rapidly with the distance away from the center Cu site. Due to the width-5 cylinder used, there are differences between the coefficient of orbitals in vertical and horizontal directions at long distance. While we do not further modify the Wannier functions, so as to preserve orthonormality, we do later average the terms over the two directions in the downfolded Hamiltonian.

\section{Further calculations of the Wannier single-band model}

After constructing of Wannier functions, we derive the downfolded single-band model by projecting the three-band Hamiltonian on the Wannier basis and truncate terms with small magnitude, as described in the main text.

\begin{figure}[h]
\centering    
\includegraphics[width=0.7\columnwidth]{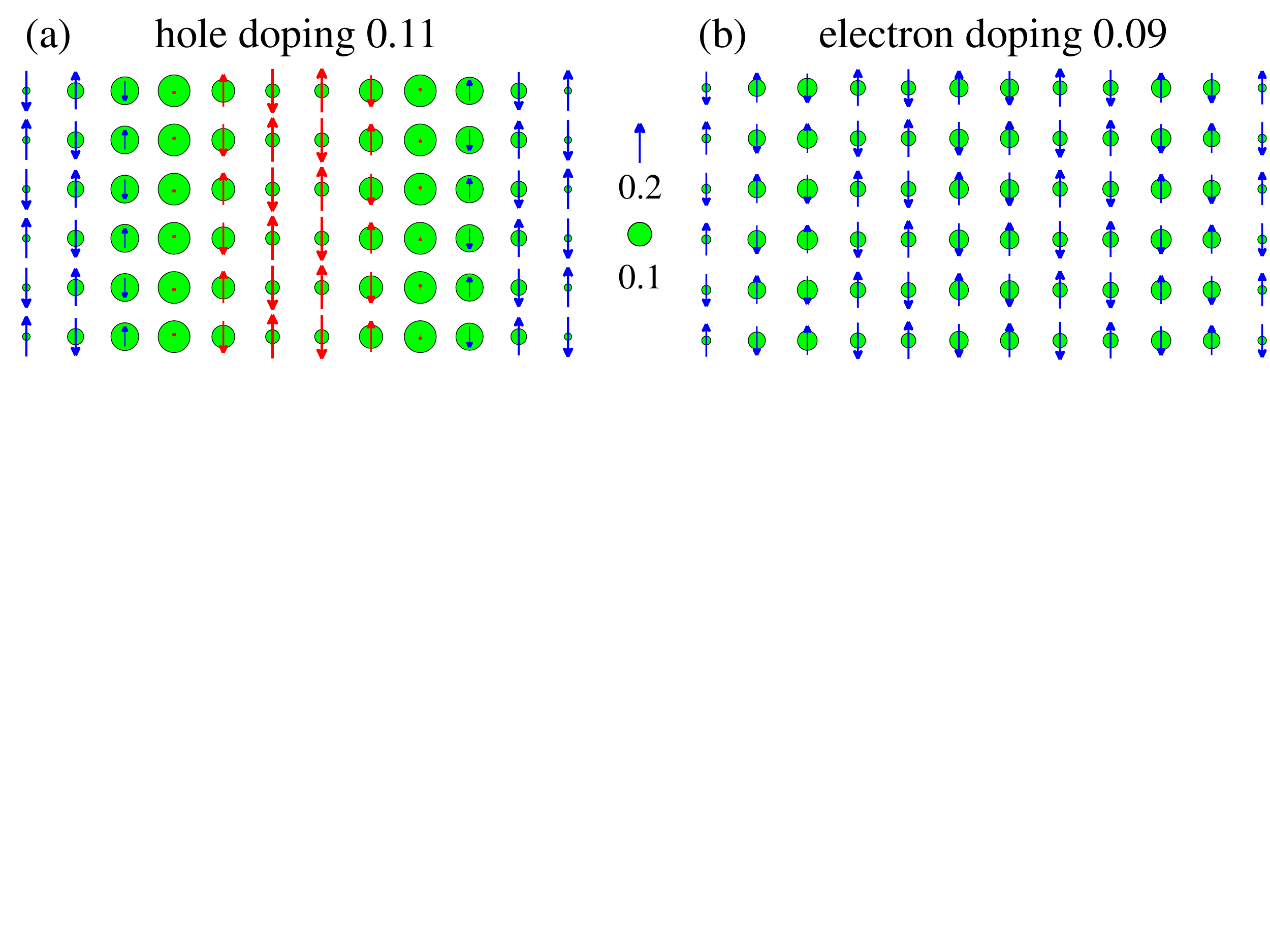}
\vskip -5.8cm
\caption{Simulations of the Wannier model on width-6 cylinders for (a) hole doping of 0.11 and (b) electron doping of 0.09, respectively.}
\label{fig:wanmod}
\end{figure}

Simulations of the Wannier model (with parameters from the case \textbf{h1}) on a width-6 cylinder for both hole and electron doping are presented in Fig.~\ref{fig:wanmod}. The occurrence of stripes (hole doping) and commensurate antiferromagnetism (electron doping) is consistent with the three-band model as well as the cuprates.

\begin{figure}[h]
\centering    
\vskip -0.5cm
\includegraphics[width=0.65\columnwidth]{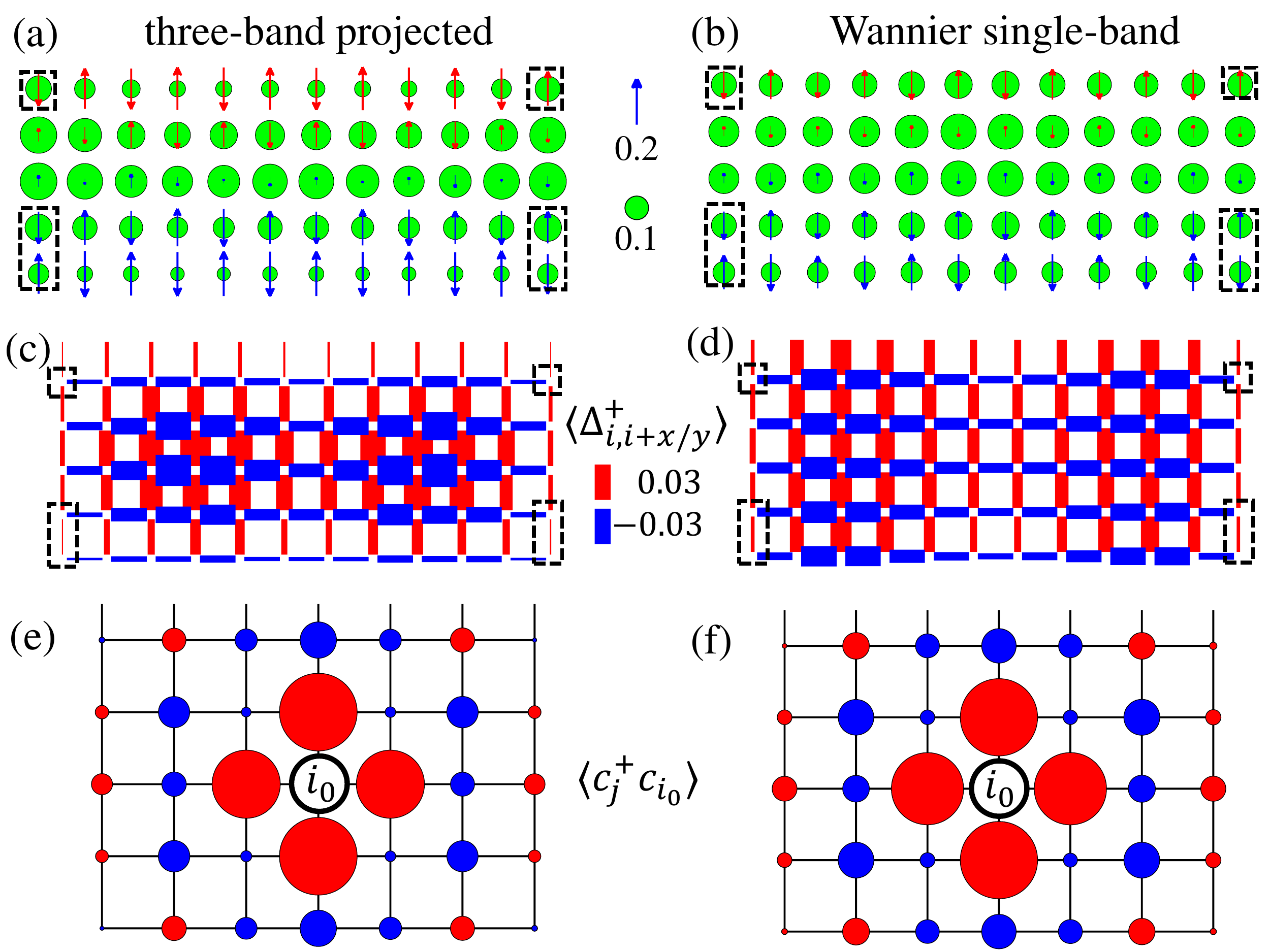}
\caption{For the case \textbf{h1}, comparisons of physical properties for the three-band wavefunction projected onto the single-band space defined by the Wannier functions (left), and a direct simulation the truncated Wannier model (right).
(a) and (b): The local charge and spin structure. The length of the arrows and the diameter of the circles represent $\langle S^z \rangle$ and local doping, respectively. 
(c) and (d): The $d$-wave pairing order $\langle \Delta^\dagger_{ij} + \Delta_{ij} \rangle$ between neighboring  sites $i$ and $j$, with the color and thickness of the bond representing the sign and amplitude of the pairing order.
(e) and (f): Single-particle correlation functions with the area and color of the circle on site $j$ indicating the magnitude and sign of $\sum_{\sigma} \langle C^\dagger_{j\sigma}C_{i_0\sigma}\rangle$, with $i_0$ being the center reference site.}
\label{fig:comparison}
\end{figure}

The validity of the Wannier model is verified by the consistency with the three-band model in the local spin and charge order, the $d$-wave pairing order, as well as the single-particle correlations, as shown in Fig.~\ref{fig:comparison}. The minor differences in modulations of the hole density and the pairing order along the $y$ direction could be related to different finite size effects or virtual processes involving higher bands, which appear to be unimportant.

\begin{figure}[ht]
\centering    
\vskip -0.5cm
\includegraphics[width=0.6\columnwidth]{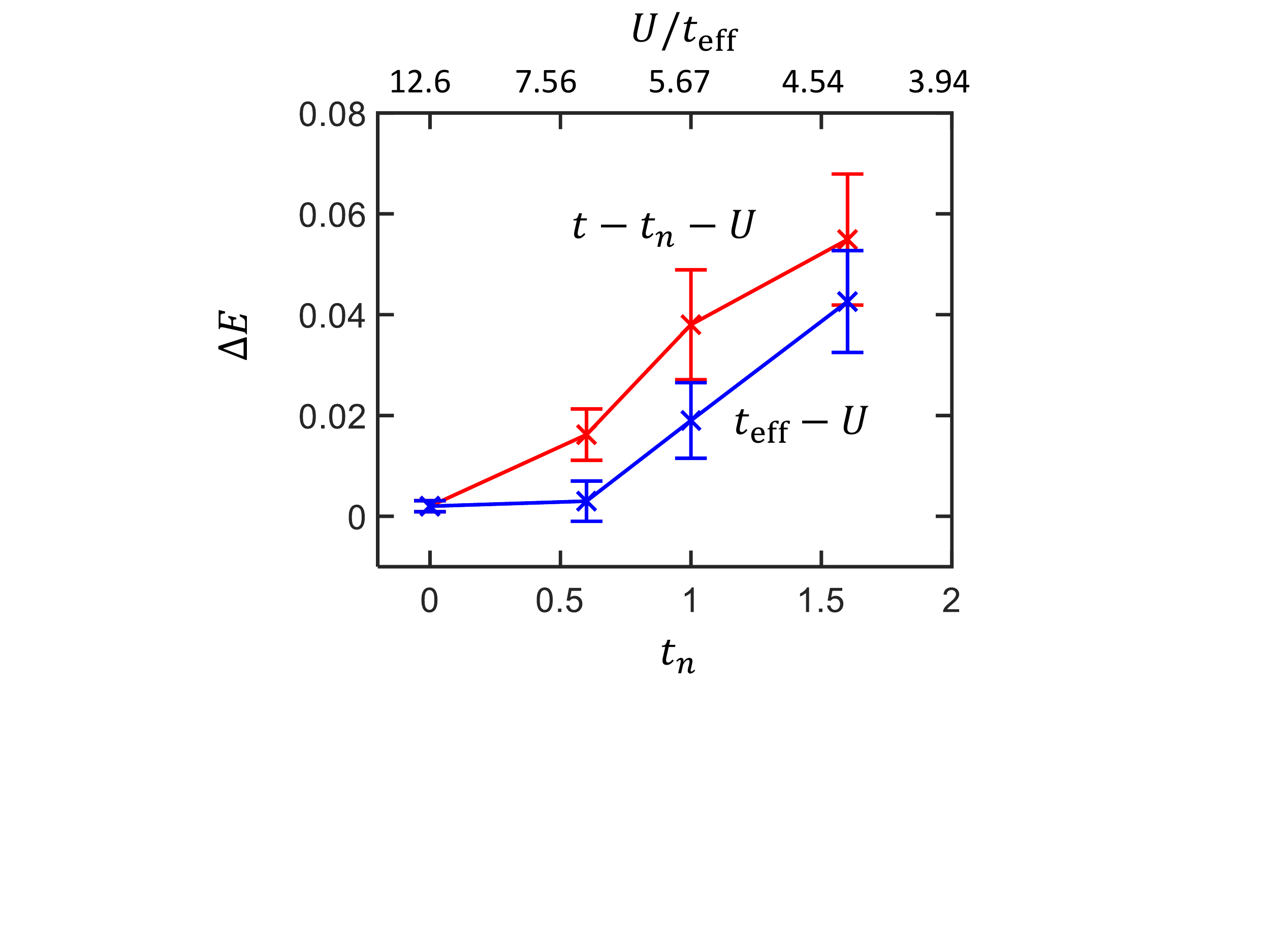}
\vskip -2.6cm
\caption{Superconducting phase stiffness measured by the energy difference for different pairfield boundary conditions (see the main text) as a function of $t_n$ and the corresponding mean-field $t_{\rm eff}$. $t=1$ is the energy unit and $U=12.6$ for both models. System is at a hole doping $\sim0.11$.}
\label{fig:phasestiff}
\end{figure}

The biggest difference between the Wannier single-band model with the conventional model is the nearest-neighbor additional density-assisted hopping term $t_n c^\dagger_{j\sigma}c_{i\sigma}(n_{i\Bar{\sigma}}+n_{j\Bar{\sigma}})$ with $t_n\sim0.6t$, which further enhances pairing comparing to its mean-field $t_nc^\dagger_{j\sigma}c_{i\sigma}\langle n\rangle$. In Fig.~\ref{fig:phasestiff}, we vary the magnitude of $t_n$ and find that it always produces a bigger superconducting phase stiffness, compared with the mean-field treatment where the $t_n$ term is incorporated into an ordinary effective hopping $t_{\rm eff}$.

\bibliography{smref}